\documentclass{article}

\usepackage{epsfig}
\usepackage{subfigure}
\usepackage{calc}
\usepackage{graphicx}

\usepackage{amsfonts}
\usepackage{amsmath}
\usepackage{amssymb}
\usepackage{latexsym}

\newtheorem{theorem}{Theorem}[section]
\newtheorem{lemma}[theorem]{Lemma}
\newtheorem{example}[theorem]{Example}
\newtheorem{remark}[theorem]{Remark}
\newtheorem{definition}[theorem]{Definition}

\begin{document}
\title{Cellular Automata in Stream Ciphers}
\date{}
\author{Amparo F\'{u}ster-Sabater \\
{\small (1) Instituto de F\'{\i}sica Aplicada, C.S.I.C.}\\
{\small Serrano 144, 28006 Madrid, Spain} \\
{\small amparo@iec.csic.es}}

\maketitle


\begin{abstract}
A wide family of nonlinear sequence generators, the so-called
clock-controlled shrinking generators, has been analyzed and
identified with a subset of linear cellular automata. The
algorithm that converts the given generator into a linear model
based on automata is very simple and can be applied in a range of
practical interest. Due to the linearity of these automata as well
as the characteristics of this class of generators, a
cryptanalytic approach can be proposed. Linear cellular structures
easily model keystream generators with application in stream
cipher cryptography.


\end{abstract}

\maketitle

\section{Introduction}
\footnotetext{Research supported by Ministerio de Educaci\'{o}n y
Ciencia
(Spain) under grant SEG2004-02418 and SEG2004-04352-C04-03. \\
Contemporary Mathematics. Volume 477, pp. 1-21, 2009.} \noindent
Cellular Automata (CA) are particular forms of finite state
machines that can be investigated by the usual analytic techniques
(\cite{Das}, \cite{Martin}, \cite{Nandi}, \cite{Wolfram1}). CA
have been used in application areas so different as physical
system simulation, biological process, species evolution,
socio-economical models or test pattern generation. They are
defined as arrays of identical cells in an \textit{n}-dimensional
space and characterized by different parameters \cite{Wolfram2}:
the cellular geometry, the neighborhood specification, the number
of contents per cell and the transition rule to compute the
successor state. Their simple, modular and cascable structure
makes them very attractive for VLSI implementations.

On the other hand, Linear Feedback Shift Registers (LFSRs)
\cite{Golomb} are linear structures currently used in the
generation of pseudorandom sequences. The inherent simplicity of
LFSRs, their ease of implementation and the good statistical
properties of their output sequences turn them into natural
building blocks for the design of pseudorandom sequence generators
with applications in spread-spectrum communications, circuit
testing, error-correcting codes, numerical simulations or
cryptography.

In recent years, one-dimensional CA have been proposed as an
alternative to LFSRs (\cite{Bao}, \cite{Blackburn}, \cite{Nandi},
 \cite{Wolfram2}) in the sense that every sequence generated by
a LFSR can be obtained from one-dimensional CA too. In
cryptographic applications, pseudorandom sequence generators
currently involve several LFSRs combined by means of nonlinear
functions or irregular clocking techniques (see \cite{Menezes},
\cite{Rueppel}). Moreover in \cite{Serra}, it is proved that
one-dimensional linear CA are isomorphic to conventional LFSRs.
Thus, the latter structures can be simply substituted by the
former ones in order to accomplish the same goal: generation of
keystream sequences.

The above class of linear CA has been found to satisfy randomness
properties with application in the testing of digital circuits,
Built-In Self-Test (BIST) schemes and self-checking \cite{Zhang}.
The current interest of these CA stems from the lack of
correlation between the bit sequences generated by adjacent cells,
see \cite{Cho}. In this sense, linear CA are superior to the more
common LFSRs \cite{Golomb} that have been traditionally used as
pseudorandom pattern generators. Nevertheless, the main advantage
of CA is that multiple generators designed as nonlinear structures
in terms of LFSRs preserve the linearity when they are expressed
under the form of CA.

This paper considers the problem of finding one-dimensional CA
that reproduce the output sequence of a particular LFSR-based
generator. More precisely, in this work a wide class of LFSR-based
nonlinear generators, the so-called Clock-Controlled Shrinking
Generators (CCSGs) \cite{Kanso}, can be described in terms of
one-dimensional CA configurations. Indeed, the well known
Shrinking Generator \cite{Coppersmith} is just an element of such
a class. The automata here presented unify in a simple structure
the above mentioned class of sequence generators. The algorithm
that converts a given CCSG into a CA-based linear model is very
simple and can be applied to CCSGs in a range of practical
interest. The underlying idea of this modelling procedure is the
concatenation of a basic automaton. Once the generators have been
linearized, a cryptanalytic approach to reconstruct the generated
sequence is also presented.

The paper is organized as follows: in section 2, the basic
structures considered, e.g. one-dimensional CA and CCSGs, are
introduced. A simple algorithm to determine the pair of CA
corresponding to a particular shrinking generator and its
generalization to Clock-Controlled Shrinking Generators are given
in sections 3 and 4, respectively. A method of reconstructing the
generated sequence that exploits the linearity of the CA-based
model is presented in section 5. Finally, conclusions in section 6
end the paper.

\section{Basic Structures}
In the following subsections, we introduce the general
characteristics of the basic structures we are dealing with:
one-dimensional cellular automata, the shrinking generator and the
class of clock-controlled shrinking generators. The work is
restricted to CA and LFSRs with binary contents. In addition, all
the LFSRs here considered are maximal-length LFSRs, that is their
output sequences are \textit{PN-}sequences and their
characteristic polynomials are primitive polynomials, see
\cite{Golomb} and \cite{Menezes}.

\subsection{One-Dimensional Cellular Automata}
One-dimensional cellular automata can be described as
\textit{L}-cell registers \cite{Cattell1}, whose cell contents are
updated at the same time instant according to a particular
\textit{k}-variable function (the \textit{transition rule})
denoted by $\Phi$. If the function $\Phi$ is a linear function, so
is the cellular automaton. In addition, for cellular automata with
binary contents there can be up to $2^{2^{k}}$ different mappings
to the next state. Moreover, if $k=2r+1$, then the binary content
of the \textit{i-th} cell at time $t+1$ depends on the contents of
$k$ neighbor cells at time $t$ in the following way:
\begin{equation}\label{eq:1}
x_{i}^{t+1}=\Phi (x_{i-r}^{t},\ldots ,x_{i}^{t},\ldots
,x_{i+r}^{t}) \;\; (i=1, ..., L).
\end{equation}

The number of cells $L$ (numbered from left to right) is the
length of the automaton. CA are called \textit{uniform} whether
all cells evolve under the same rule while CA are called
\textit{hybrid} whether different cells evolve under different
rules. At the ends of the array, two different boundary conditions
are possible: \textit{null automata} when cells with permanent
null contents are supposed adjacent to the extreme cells or
\textit{periodic automata} when extreme cells are supposed
adjacent.

In this paper, only transition rules with $k=3$ will be
considered. Thus, there are $2^8$ of such rules among which just
two (rule $90$ and rule $150$) lead to non trivial machines. Such
rules are described as follows :

\begin{center}
Rule 90\\
$x_{i}^{t+1}=\Phi_{90}(x_{i-1}^{t},x_{i}^{t},x_{i+1}^{t})=x_{i-1}^{t}+x_{i+1}^{t}$\\

\begin{center}
$
\begin{array}{cccccccc}
111 & 110 & 101 & 100 & 011 & 010 & 001 & 000 \\
0 & 1 & 0 & 1 & 1 & 0 & 1 & 0 \\
\end{array}
$

\end{center}
\end{center}

\vspace*{0.2cm}
\begin{center}
Rule 150\\
$x_{i}^{t+1}=\Phi_{150}(x_{i-1}^{t},x_{i}^{t},x_{i+1}^{t})=x_{i-1}^{t}+x_{i}^{t}+x_{i+1}^{t}$\\

\begin{center}
$
\begin{array}{cccccccc}
111 & 110 & 101 & 100 & 011 & 010 & 001 & 000 \\
1 & 0 & 0 & 1 & 0 & 1 & 1 & 0 \\
\end{array}
$
\end{center}
\end{center}
\vspace*{0.2cm}

Remark that the names rule 90 and rule 150 derive from the decimal
values of their next-state functions: $01011010$ (binary) = $90$
(decimal) and $10010110$ (binary) = $150$ (decimal). Indeed,
$x_{i}^{t+1}$ the content of the \textit{i-th} cell at time $t+1$
depends on the contents of either two different cells (rule 90) or
three different cells (rule 150) at time $t$. The symbol $+$
denotes addition modulo $2$ among cell contents. Remark that both
transition rules are linear. This work deals exclusively with
one-dimensional linear null hybrid CA with rules 90 and 150. A
natural way of specifying such CA is an \textit{L-}tuple $M=[R_1,
R_2, ..., R_L]$, called \textit{rule vector}, where $R_i=0$ if the
\textit{i-th} cell satisfies rule 90 while $R_i=1$ if the
\textit{i-th} cell satisfies rule 150. A sub-automaton of the
previous automata class consisting of cells 1 through $i$ will be
denoted by $R_1R_2...R_i$.

\begin{table}[ht]
\caption{An one-dimensional linear null hybrid cellular automaton
of $10$ cells with rules 90/150 starting at a given initial state}
\label{table:1}
\noindent\[
\begin{tabular}{cccccccccc}
\hline\noalign{\smallskip}
$\;\;90\;$ & $\;\;150$ & $\;\;150$ & $\;\;150$ & $\;\;90\;$ & $\;\;90\;$ & $\;\;150$ & $\;\;150$ & $\;\;150$ & $\;\;90\;$ \\
\noalign{\smallskip} \hline \noalign{\smallskip} \hline
$\;0$ & $\;0$ & 0 & 1 & 1 & 1 & 0 & 1 & 1 & 0 \\
$\;0$ & $\;0$ & 1 & 0 & 0 & 1 & 0 & 0 & 0 & 1 \\
$\;0$ & $\;1$ & 1 & 1 & 1 & 0 & 1 & 0 & 1 & 0 \\
$\;1$ & $\;0$ & 1 & 1 & 1 & 0 & 1 & 0 & 1 & 1 \\
$\;0$ & $\;0$ & 0 & 1 & 1 & 0 & 1 & 0 & 0 & 1 \\
$\;0$ & $\;0$ & 1 & 0 & 1 & 0 & 1 & 1 & 1 & 0 \\
$\;0$ & $\;1$ & 1 & 0 & 0 & 0 & 0 & 1 & 0 & 1 \\
$\;1$ & $\;0$ & 0 & 1 & 0 & 0 & 1 & 1 & 0 & 0 \\
$\;0$ & $\;1$ & 1 & 1 & 1 & 1 & 0 & 0 & 1 & 0 \\
$\;1$ & $\;0$ & 1 & 1 & 0 & 1 & 1 & 1 & 1 & 1 \\
$\;\vdots$ & $\;\vdots$ & \vdots & \vdots & \vdots & \vdots & \vdots & \vdots & \vdots & \vdots \\
\hline
\end{tabular}
\]
\end{table}

For a cellular automaton of length $L=10$ cells, configuration
rules
$(\,R_1=0,R_2=1,R_3=1,R_4=1,R_5=0,R_6=0,R_7=1,R_8=1,R_9=1,R_{10}=0\,)$
and initial state $(0,0,0,1,1,1,0,1,1,0)$, Table \ref{table:1}
illustrates the formation of its output sequences (binary
sequences read vertically) and the succession of states (binary
configurations of 10 bits read horizontally). For the above
mentioned rules, the different states of the automaton are grouped
in closed cycles. The number of different output sequences for a
particular cycle is $\leq L$ as the same sequence (although
shifted) may appear simultaneously in different cells. At the same
time, all the sequences in a cycle will have the same period and
linear complexity \cite{Martin}. Moreover, any of the output
sequence of the automaton can be produced at any cell provided
that the right state cycle is chosen.

On the other hand, linear finite state machines are currently
represented and analyzed by means of their transition matrices.
The form and characteristics of these matrices for the CA under
consideration can be found in \cite{Cattell1}. In fact, such
matrices are tri-diagonal matrices with the rule vector on the
main diagonal, 1's on the diagonals below and above the main one
and all other entries being zero. Every automaton is completely
specified by its characteristic polynomial, that is the
characteristic polynomial of its transition matrix. Such a
characteristic polynomial can be computed in terms of the
characteristic polynomials of the previous sub-automata according
to the recurrence relation \cite{Cattell1}:
\begin{equation}\label{equation:0}
P_{i}(x)=(x+R_i)P_{i-1}(x)+P_{i-2}(x), \;\; 0<i\leq L
\end{equation}
being $P_{-1}(x)=0$ and $P_{0}(x)=1$. Next, the following
definition is introduced:
\begin{definition}\label{definition:1}
A Multiplicative-Polynomial Cellular Automaton is defined as a
cellular automaton whose characteristic polynomial is a reducible
polynomial of the form $P_M(x)= (P(x))^{p}$ where $p$ is a
positive integer. If $P(x)$ is a primitive polynomial, then the
automaton is called a Primitive Multiplicative-Polynomial Cellular
Automaton.
\end{definition}
The class of binary sequence generators we are dealing with is
described in the following subsections.
\subsection{The Shrinking Generator}\label{subsection:2}
The shrinking generator is a binary sequence generator
\cite{Coppersmith} composed by two LFSRs : a control register
$SR_{1}$ that decimates the sequence produced by the other
register $SR_{2}$. We denote by $L_j\; (j=1,2)$ their
corresponding lengths with $(L_1, L_2)=1$ as well as  $L_1<L_2$.
Then, we denote by $C_j(x)\in GF(2)[x]\; (j=1,2)$ their
corresponding characteristic polynomials of degree $L_j\;
(j=1,2)$, respectively.

The sequence produced by $SR_{1}$, denoted by $\{a_{i}\}$,
controls the bits of the sequence produced by $SR_{2}$, that is
$\{b_{i}\}$, which are included in the output sequence $\{z_{j}\}$
(the \textit{shrunken sequence}), according to the following rule
$P$:

\begin{enumerate}
\item  If $a_{i}=1\Longrightarrow z_{j}=b_{i}$

\item  If $a_{i}=0\Longrightarrow b_{i}$ is discarded.
\end{enumerate}

A simple example illustrates the behavior of this structure.

\begin{example} Let us consider the following LFSRs:

\begin{enumerate}
\item  Register $SR_{1}$ of length $L_{1}=3$, characteristic
polynomial $C_{1}(x)=1+x^{2}+x^{3}$ and initial state
$IS_{1}=(1,0,0)$. The \textit{PN-}sequence generated by $SR_{1}$
is $\{1,0,0,1,1,1,0\}$ with period $T_1=2^{L_1}-1=7$.

\item  Register $SR_{2}$ of length $L_{2}=4$, characteristic
polynomial $C_{2}(x)=1+x+x^{4}$ and initial state
$IS_{2}=(1,0,0,0)$. \thinspace The \textit{PN-}sequence generated
by $SR_{2}$ is $\{1,0,0,0,1,0,0,1,1,$ $0,1,0,1,1,1\}$ with period
$T_2=2^{L_2}-1=15$.
\end{enumerate}

The output sequence $\{z_{j}\}$ is given by:
\begin{itemize}
\item  $\{a_{i}\}$ $\rightarrow $ $1\;0\;0\;1\;1\;1\;0\;1\;0\;0\;1\;1\;1\;0%
\;1\;0\;0\;1\;1\;1\;0\;1\;.....$

\item  $\{b_{i}\}$ $\rightarrow $ $\hspace{0.02cm}1\;\underline{0}\;\underline{0}\;0\;1\;0\;%
\underline{0}\;1\;\underline{1}\;\underline{0}\;1\;0\;1\;\underline{1}\;1\;\underline{1}%
        \;\underline{0}\;0\;0\;1\;\underline{0}\;0\;.....$

\item  $\{z_{j}\}$ $\rightarrow $
$1\;0\;1\;0\;1\;1\;0\;1\;1\;0\;0\;1\;0\;.....$
\end{itemize}
The underlined
bits \underline{0} or \underline{1} in $\{b_{i}\}$ are discarded.

\end{example}

In brief, the sequence produced by the shrinking generator is an
irregular decimation of $\{b_{i}\}$ from the bits of $\{a_{i}\}$.
According to \cite{Coppersmith}, the period of the shrunken
sequence is
\begin{equation}\label{equation:1}
T=(2^{L_{2}}-1)2^{(L_{1}-1)}
\end{equation}
and its linear complexity \cite{Rueppel}, notated $LC$, satisfies
the following inequality
\begin{equation}\label{equation:2}
L_{2} \thinspace 2^{(L_{1}-2)}<LC\leq L_{2} \thinspace
2^{(L_{1}-1)}.
\end{equation}
A simple calculation, based on the fact that every state of
$SR_{2}$ coincides once with every state of $SR_{1}$, allows one
to compute the number of $1$'s in the shrunken sequence. Such a
number is constant and equal to
\begin{equation}\label{equation:3}
No. \thickspace 1's = 2^{(L_{2}-1)}2^{(L_{1}-1)}.
\end{equation}
Comparing period and number of $1's$, it can be concluded that the
shrunken sequence is a quasi-balanced sequence.

In addition, it can be proved \cite{Coppersmith} that the output
sequence has good distributional statistics too. Therefore, this
scheme is suitable for practical implementation of stream ciphers
and pattern generators.

\subsection{The Clock-Controlled Shrinking Generators}
The Clock-Controlled Shrinking Generators constitute a wide class
of clock-controlled sequence generators \cite{Kanso} with
applications in cryptography, error correcting codes and digital
signature. An CCSG is a sequence generator composed of two LFSRs
notated $SR_{1}$ and $SR_{2}$. The parameters of both registers
are defined as those of subsection \ref{subsection:2}. At any time
$t$, $SR_{1}$ (the control register) is clocked normally while the
second register $SR_{2}$ is clocked a number of times given by an
integer decimation function notated $X_t$. In fact, if $A_0(t),\,
A_1(t),\,\ldots,\, A_{L_{1}-1}(t)$ are the binary cell contents of
$SR_{1}$ at time $t$, then $X_t$ is defined as
\begin{equation}\label{equation:4}
X_{t}=1+2^0 A_{i_0}(t) + 2^1 A_{i_1}(t) + \ldots + 2^{w-1}
A_{i_{w-1}}(t)
\end{equation}
where $i_0,\,i_1,\,\ldots,\, i_{w-1}\in \{0,\, 1,\, \ldots,
\,L_{1}-1\}$ and $0 < w \leq L_{1}-1$.

In this way, the output sequence of an CCSG is obtained from a
double decimation:
\begin{enumerate}
\item The output sequence of $SR_{2}$, $\{b_{i}\}$, is decimated
by means of $X_t$ giving rise to the sequence $\{b'_{i}\}$. \item
The same decimation rule $P$, defined in subsection
\ref{subsection:2}, is applied to the sequence $\{b'_{i}\}$.
\end{enumerate}
Remark that if $X_t \equiv 1$ (no cells are selected in $SR_1$),
then the proposed generator is just the shrinking generator. Let
us see a simple example of CCSG.

\begin{example} For the same LFSRs defined in the previous
example and the function $X_{t}=1+2^0 A_{0}(t)$ with $w=1$, the
decimated sequence $\{b'_{i}\}$ is given by:

\begin{itemize}
\item  $\{b_{i}\}$ $\rightarrow $ $1\;\underline{0}\;0\;0\;1\;\underline{0}\;0\;\underline{1}\;1\;\underline{0}\;1\;0\;\underline{1}\;1\;1%
\;1\;\underline{0}\;0\;\underline{0}\;1\;\underline{0}\;0\;1\;\underline{1}\;0\;1\;0\;\underline{1}\;1\;\underline{1}\;1\;.....$

\item  $\;X_{t}\;$ $\rightarrow $
$2\;1\;1\;2\;2\;2\;1\;2\;1\;1\;2\;2\;2\;1\;2\;1\;1\;2\;2\;.....$

\item  $\{b'_{i}\}$ $\rightarrow $ $1\;0\;0\;1\;0\;1\;%
1\;0\;1\;1\;1\;0\;1\;0\;1\;0%
        \;1\;0\;1\;1\;.....$
\end{itemize}
According to the decimation function $X_t$, the underlined bits
\underline{0} or \underline{1} in $\{b_{i}\}$ are discarded in
order to produce the sequence $\{b'_{i}\}$. Then the output
sequence $\{z_{j}\}$ of the CCSG is given by:

\begin{itemize}
\item  $\{a_{i}\}$ $\rightarrow $ $1\;0\;0\;1\;1\;1\;0\;1\;0\;0\;1\;1\;1\;0%
\;1\;0\;0\;1\;1\;1\;0\;1\;.....$

\item  $\{b'_{i}\}$ $\rightarrow $ $\hspace{0.02cm}1\;\underline{0}\;\underline{0}\;1\;0\;1\;%
\underline{1}\;0\;\underline{1}\;\underline{1}\;1\;0\;1\;\underline{0}\;1\;\underline{0}%
        \;\underline{1}\;0\;1\;1\;.....$

\item  $\{z_{j}\}$ $\rightarrow $
$1\;1\;0\;1\;0\;1\;0\;1\;1\;0\;1\;1\;.....$ \\
\end{itemize}
The underlined bits \underline{0} or \underline{1} in $\{b'_{i}\}$
are discarded.
\end{example}

In brief, the sequence produced by an CCSG is an irregular double
decimation of the sequence generated by $SR_2$ from the function
$X_t$ and the bits of $SR_1$. This construction allows one to
generate a large family of different sequences by using the same
LFSR initial states and characteristic polynomials but modifying
the decimation function. Period, linear complexity and statistical
properties of the generated sequences by CCSGs have been
established in \cite{Kanso}.

\subsection{Cattel and Muzio Synthesis Algorithm}
The Cattell and Muzio synthesis algorithm \cite{Cattell2} presents
a method of obtaining two CA (based on rules 90 and 150)
corresponding to a given polynomial. Such an algorithm takes as
input an irreducible polynomial $Q(x) \in GF(2)[x]$ defined over a
finite field and computes two linear reversal CA whose output
sequences have $Q(x)$ as characteristic polynomial. Such CA are
written as binary strings with the previous codification: $0$ =
rule $90$ and $1$ = rule $150$. The theoretical foundations of the
algorithm can be found in \cite{Cattell4}. The total number of
operations required for this algorithm is listed in
\cite{Cattell2}(Table II, page 334). It is shown that the number
of operations grows linearly with the degree of the polynomial, so
the method does not suffer from any sort of exponential blow-up.
The method is efficient for all practical applications (e.g. in
1996 finding a pair of length $300$ CA took 16 CPU seconds on a
SPARC 10 workstation). For cryptographic applications, the degree
of the irreducible (primitive) polynomial is $ L_{2}\approx 64$,
so that the consuming time is negligible.

Finally, a list of One-Dimensional Linear Hybrid Cellular Automata
of Degree Through 500 can be found in \cite{Cattell3}.

\section{CA-Based Linear Models for the Shrinking Generator}\label{section:3}

In this section, an algorithm to determine the pair of CA
corresponding to a given shrinking generator is presented. Such an
algorithm is based on the following results:

\begin{lemma}\label{lemma:1}
The characteristic polynomial of the shrunken sequence is of the
form $P(x)^{N}$, where $P(x)\in GF(2)[x]$ is a $L_{2}$-degree
primitive polynomial and $N$ is an integer satisfying the
inequality $2^{(L_{1}-2)}<N\leq 2^{(L_{1}-1)}$.
\end{lemma}
\textbf{Proof:} The shrunken sequence can be written as an
interleaved sequence \cite{Gong} made out of an unique
\textit{PN}-sequence starting at different points and repeated
$2^{(L_{1}-1)}$ times. Such a sequence is obtained from
$\{b_{i}\}$ taking digits separated a distance $2^{L_1}-1$, that
is the period of the sequence $\{a_{i}\}$. As $(2^{L_2}-1,
2^{L_1}-1)=1$ due to the primality of $L_2$ and $L_1$, the result
of the decimation of $\{b_{i}\}$ is a \textit{PN}-sequence of
primitive characteristic polynomial $P(x)$ of degree $L_2$.
Moreover, the number of times that this \textit{PN}-sequence is
repeated coincides with the number of $1's$ in $\{a_{i}\}$ since
each $1$ of $\{a_{i}\}$ provides the shrunken sequence with
$2^{L_2}-1$ digits of $\{b_{i}\}$. Consequently, the
characteristic polynomial of the shrunken sequence will be
$P(x)^{N}$ with $N\leq 2^{(L_{1}-1)}$. The lower limit follows
immediately from equation (\ref{equation:2}) that defines the
linear recurrence relationship.\hfill $\Box$

\begin{lemma}\label{lemma:2}
Let $C_{2}(x)\in GF(2)[x]$ be the characteristic polynomial of
$SR_2$ and let $\lambda$ be a root of $C_{2}(x)$ in the extension
field $GF(2^{L_{2}})$. Then, $P(x)\in GF(2)[x]$ is of the form

\begin{equation}\label{equation:5}
P(x) =(x+\lambda^{E})(x+\lambda^{2E})\ldots
(x+\lambda^{2^{L_{2}-1}E})
\end{equation}
being $E$ an integer given by
\begin{equation}\label{equation:6}
E =2^{0}+2^{1}+\ldots +2^{L_{1}-1} \;.
\end{equation}
\end{lemma}
\textbf{Proof:} As the decimation of the sequence $\{b_{i}\}$ is
realized taking one out of $2^{L_1}-1$ digits, the obtained
\textit{PN}-sequence is nothing but the characteristic sequence
associated to the \textit{cyclotomic coset} $E=2^{L_{1}}-1$, see
\cite{Golomb}. Hence, the roots of its characteristic polynomial
will be $\lambda^{E}, \lambda^{2E}, \dots,
\lambda^{2^{L_{2}-1}E}$. According to the definition of cyclotomic
coset, the value of $E$ is given by equation
(\ref{equation:6}).\hfill $\Box$

Remark that $P(x)$ depends exclusively on the characteristic
polynomial of the register $SR_{2}$ and on the length $L_{1}$ of
the register $SR_{1}$. Based on the Cattell and Muzio synthesis
algorithm \cite{Cattell2}, the following result is derived:

\begin{lemma}\label{lemma:3}
Let $Q(x)\in GF(2)[x]$ be a polynomial defined over a finite field
and let $s_1$ and $s_2$ two binary strings codifying the two
linear CA obtained from the Cattell and Muzio algorithm. Then, the
two CA in form of binary strings whose characteristic polynomial
is $Q(x)^2$ are:
 \[
  S'_{i} = {S_i}*{S_i^{*}}\;\;\; i=1, 2
  \]
where $S_i$ is the binary string $s_i$ whose least significant bit
has been complemented, $S_i^{*}$ is the mirror image of $S_i$ and
the symbol $*$ denotes concatenation.
\end{lemma}
\textbf{Proof:} The result is just a generalization of the Cattell
and Muzio synthesis algorithm. The concatenation is due to the
fact that rule $90$ ($150$) at the end of the array in null
automata is equivalent to two consecutive rules $150$ ($90$) with
identical sequences. The fact of that an automaton and its
reversal version have the same characteristic polynomial completes
the proof.\hfill $\Box$

Proceeding in the same way a number of times, a
multiplicative-polynomial cellular automaton \ref{definition:1} is
obtained. In this way, the construction of a linear structure from
the concatenation of a basic automaton is accomplished.

According to the previous results, an algorithm to linearize the
shrinking generator is introduced:

\bigskip
\textbf{Input:} A shrinking generator characterized by two LFSRs,
$SR_{1}$ and $SR_{2}$, with their corresponding lengths, $L_{1}$
and $L_{2}$, and the characteristic polynomial $C_{2}(x)$ of the
register $SR_{2}$.

\begin{description}
\item [Step 1] From $L_{1}$ and $C_{2}(x)$, compute the polynomial
$P(x)$ in $GF(2^{L_{2}})$ as
\begin{equation*}
P(x) =(x+\lambda^{E})(x+\lambda^{2E})\ldots
(x+\lambda^{2^{L_{2}-1}E})
\end{equation*}
with $E =2^{0}+2^{1}+\ldots +2^{L_{1}-1}$.

\item [Step 2] From $P(x)$, apply the Cattell and Muzio synthesis
algorithm to determine two linear CA (with rules 90 and 150),
notated $s_i$, whose characteristic polynomial is $P(x)$.

\item [Step 3] For each $s_i$ separately, proceed:
\begin{description}
  \item [3.1] Complement its least significant bit. The resulting binary string is notated $S_i$.

  \item [3.2] Compute the mirror image of $S_i$, notated $S_i^{*}$, and
  concatenate both strings
  \[
  S'_{i} = S_i*S_i^{*}\;.
  \]
  \item [3.3] Apply steps $3.1$ and $3.2$ to each $S'_{i}$ recursively $L_{1}-1$ times.
\end{description}

\end{description}

\textbf{Output: } Two binary strings of length $L=L_2 \cdot
2^{L_{1}-1}$ codifying two CA corresponding to the given shrinking
generator.

\begin{remark}
In this algorithm the characteristic polynomial of the register
$SR_{1}$ is not needed. Thus, all the shrinking generators with
the same $SR_2$ but different registers $SR_1$ (all of them with
the same length $L_1$) can be modelled by the same pair of
one-dimensional linear CA.
\end{remark}
\begin{remark}
It can be noticed that the computation of both CA is proportional
to $L_{1}$ concatenations. Consequently, the algorithm can be
applied to shrinking generators in a range of practical
application.
\end{remark}
\begin{remark}
In contrast to the nonlinearity of the shrinking generator, the
CA-based models that generate the shrunken sequence are linear.
\end{remark}

In order to clarify the previous steps a simple numerical example
is presented.

\bigskip
\textbf{Input:} A shrinking generator characterized by two LFSRs
$SR_{1}$ of length $L_{1}=3$ and $SR_{2}$ of length $L_{2}=5$ and
characteristic polynomial $C_{2}(x)=1+x+x^{2}+x^{4}+x^{5}$.

\begin{description}
\item [Step 1] $P(x)$ is the characteristic polynomial of the
cyclotomic \textit{coset} $E=7$. Thus,
\[
P(x) =1+x^{2}+x^{5}\;.
\]

\item [Step 2] From $P(x)$ and applying the Cattell and Muzio
synthesis algorithm, two reversal linear CA whose characteristic
polynomial is $P(x)$ can be determined. Such CA are written in
binary format as:
\begin{center}
$
\begin{array}{ccccc}
0 & 1 & 1 & 1 & 1 \\
1 & 1 & 1 & 1 & 0
\end{array}
$
\end{center}

\item [Step 3] Computation of the required pair of CA.\\
For the first automaton:
\begin{center}
$
\begin{array}{cccccccccccccccccccc}
0 & 1 & 1 & 1 & 1 \\
0 & 1 & 1 & 1 & 0 & 0 & 1 & 1 & 1 & 0\\
0 & 1 & 1 & 1 & 0 & 0 & 1 & 1 & 1 & 1 & 1 & 1 & 1 & 1 & 0 & 0 & 1
& 1 & 1 & 0 \\
\end{array}
$
\end{center}
For the second automaton:
\begin{center}
$
\begin{array}{cccccccccccccccccccc}
1 & 1 & 1 & 1 & 0 \\
1 & 1 & 1 & 1 & 1 & 1 & 1 & 1 & 1 & 1\\
1 & 1 & 1 & 1 & 1 & 1 & 1 & 1 & 1 & 0 & 0 & 1 & 1 & 1 & 1 & 1 & 1
& 1 & 1 & 1
\end{array}
$
\end{center}
For each automaton, the procedure of concatenation has been
carried out $L_{1}-1$ times.
\end{description}

\textbf{Output: } Two binary strings of length $L = L_2 \cdot
2^{(L_{1}-1)}=20$ codifying the required pair of CA.

In this way, we have obtained a pair of linear CA able to generate
the shrunken sequence corresponding to the given shrinking
generator. In addition, for each one of the previous automata
there is one state cycle where the shrunken sequence is generated
at each one of the cells.

\section{CA-Based Linear Models for the Clock-Controlled Shrinking Generators}

In this section, an algorithm to determine the pair of
one-dimensional linear CA corresponding to a given CCSG is
presented. Such an algorithm is based on the following results:

\begin{lemma}\label{lemma:4}
 The characteristic polynomial of the output sequence of a CCSG
is of the form $P'(x)^{N}$, where $P'(x)\in GF(2)[x]$ is a
primitive $L_{2}$-degree polynomial and $N$ is an integer
satisfying the inequality $2^{(L_{1}-2)}<N\leq 2^{(L_{1}-1)}$.
\end{lemma}
\textbf{Proof:} The proof is analogous to that one developed in
lemma \ref{lemma:1}.\hfill $\Box$

Remark that, according to the structure of the CCSGs, the
polynomial $P'(x)$ depends on the characteristic polynomial of the
register $SR_{2}$, the length $L_{1}$ of the register $SR_{1}$ and
the decimation function $X_t$. Before, $P(x)$ was the
characteristic polynomial of the \textit{cyclotomic coset} $E$,
where $E=2^{0}+2^{1}+\ldots +2^{L_{1}-1}$ was a fixed separation
distance between the digits drawn from the sequence $\{b_{i}\}$.
Now, this distance $D$ is variable as well as a function of $X_t$.
The computation of $D$ gives rise to the following result:

\begin{lemma}\label{lemma:5}
Let $C_{2}(x)\in GF(2)[x]$ be the characteristic polynomial of
$SR_2$ and let $\lambda$ be a root of $C_{2}(x)$ in the extension
field $GF(2^{L_{2}})$. Then, $P'(x)\in GF(2)[x]$ is the
characteristic polynomial of \textit{cyclotomic coset} $D$, where
$D$ is given by
\begin{equation}\label{equation:7}
D=2^{L_1-w}\; (\sum \limits_{i=1}^{2^w}i) \;
-1=(1+2^w)\;2^{L_1-1}\;-1.
\end{equation}
\end{lemma}

\textbf{Proof:} The proof is analogous to that one developed in
lemma \ref{lemma:2}. In fact, the distance $D$ can be computed
taking into account that the function $X_t$ takes values in the
interval $[1,\;2,\;\ldots,\;2^w]$ and the number of times that
each one of these values appears in a period of the output
sequence is given by $2^{L_1-w}$. A simple computation, based on
the sum of the terms of an arithmetic progression, completes the
proof.\hfill $\Box$

From the previous results, it can be noticed that the algorithm
that determines the pair of CA corresponding to a given CCSG is
analogous to that one developed in section \ref{section:3}.
Indeed, the expression of $E$ in equation (\ref{equation:6}) must
be replaced by the expression of $D$ in equation
(\ref{equation:7}).

In order to clarify the previous steps a simple numerical example
is presented.

\bigskip
\textbf{Input:} A CCSG characterized by: Two LFSRs $SR_{1}$ of
length $L_{1}=3$ and $SR_{2}$ of length $L_{2}=5$ and
characteristic polynomial $C_{2}(x)=1+x+x^{2}+x^{4}+x^{5}$ plus
the decimation function $X_{t}=1+2^0 A_{0}(t) + 2^1 A_{1}(t) + 2^2
A_{2}(t)$ with $w=3$.

\begin{description}
\item [Step 1] $P'(x)$ is the characteristic polynomial of the
cyclotomic \textit{coset D}. Now $D \equiv 4\;mod\;31$, that is we
are dealing with the cyclotomic coset $1$. Thus, the corresponding
characteristic polynomial is:
\[
P'(x) =1+x+x^{2}+x^{4}+x^{5}\;.
\]

\item [Step 2] From $P'(x)$ and applying the Cattell and Muzio
synthesis algorithm, two reversal linear CA whose characteristic
polynomial is $P'(x)$ can be determined. Such CA are written in
binary format as:
\begin{center}
$
\begin{array}{ccccc}
1 & 0 & 0 & 0 & 0 \\
0 & 0 & 0 & 0 & 1
\end{array}
$
\end{center}

\item [Step 3] Computation of the required pair of CA.\\
For the first automaton:
\begin{center}
$
\begin{array}{cccccccccccccccccccc}
1 & 0 & 0 & 0 & 0 \\
1 & 0 & 0 & 0 & 1 & 1 & 0 & 0 & 0 & 1\\
1 & 0 & 0 & 0 & 1 & 1 & 0 & 0 & 0 & 0 & 0 & 0 & 0 & 0 & 1 & 1 & 0
& 0 & 0 & 1 \\
\end{array}
$
\end{center}
For the second automaton:
\begin{center}
$
\begin{array}{cccccccccccccccccccc}
0 & 0 & 0 & 0 & 1 \\
0 & 0 & 0 & 0 & 0 & 0 & 0 & 0 & 0 & 0\\
0 & 0 & 0 & 0 & 0 & 0 & 0 & 0 & 0 & 1 & 1 & 0 & 0 & 0 & 0 & 0 & 0
& 0 & 0 & 0
\end{array}
$
\end{center}
For each automaton, the procedure of concatenation has been
carried out $L_{1}-1$ times.
\end{description}

\textbf{Output: } Two binary strings of length $L = 20$ codifying
the required CA.

\begin{remark}
From a point of view of the CA-based linear models, the shrinking
generator or any one of the CCGS are entirely analogous. Thus, the
fact of introduce an additional decimation function does neither
increase the complexity of the generator nor improve its
resistance against cryptanalytic attacks. Indeed, both kinds of
generators can be linearized by the same class of CA-based models.
\end{remark}

\section{A Cryptanalytic Approach to this Class of Sequence Generators}

Since CA-based linear models describing the behavior of CCSGs have
been derived, a cryptanalytic attack that exploits the weaknesses
of these models has been also developed. It consists in
determining the initial states of both registers $SR_1$ and $SR_2$
from an amount of CCSG output sequence (the \textit{intercepted
sequence}). In this way, the rest of the output sequence can be
reconstructed. For the sake of simplicity, the attack will be
illustrated for the shrinking generator although the process can
be extended to any CCSG. The proposed attack is divided into two
different phases:
\begin{description}
\item [Phase 1] From bits of the intercepted sequence and using
the CA-based linear models, additional bits of the shrunken
sequence can be reconstructed. \item [Phase 2] Due to the
intrinsic characteristics of this class of generators, a
cryptanalytic attack can be mounted in order to determine the
initial states of the LFSRs. The attack makes use of both
intercepted bits as well as reconstructed bits.
\end{description}
Both phases will be considered separately.
\subsection{Reconstruction of output sequence bits}
Given $r$ bits of the shrunken sequence $z_0,z_1,z_2,...,z_{r-1}$
, we can assume without loss of generality that this sub-sequence
has been generated at the most left extreme cell of any of its
corresponding CA. That is $x_{1}^{t}=z_0,\; x_{1}^{t+1}=z_1, \;
...,\; x_{1}^{t+r-1}=z_{r-1}$. From $r$ bits of the shrunken
sequence, it is always possible to reconstruct $r-1$ sub-sequences
$\{x_{i}^{t}\}$ of lengths $r-i+1$ at the \textit{i-th} cell of
each automaton such as follows:
\begin{equation}\label{equation:8}
x_{i}^{t}=\Phi_{i-1}(x_{i-2}^{t},x_{i-1}^{t},x_{i-1}^{t+1}) \;\;\;
(1<i \leq r),
\end{equation}
where $\Phi_{i-1}$ corresponds to either rule 90 or 150 depending
on the value of $R_{i-1}$. From $r$ intercepted bits, the
application of equation (\ref{equation:8}) gives rise to a total
of $(r+(r-1)+\ldots+2+1)$ bits that constitute the first chained
sub-triangle notated $\Delta 1$, see Table \ref{table:headings2}.
Now, if any sub-sequence $\{x_{i}^{t}\}$ is placed at the most
left extreme cell, then $r-2i+2$ bits are obtained at the
\textit{i-th} cell in the second chained sub-triangle notated
$\Delta 2$. Repeating recursively $n$ times the same procedure,
$r-ni+n$ bits are obtained at the \textit{i-th} cell in the
\textit{n-th} chained sub-triangle notated $\Delta n$. Table
\ref{table:headings2} shows the succession of 4 chained
sub-triangles constructed from $r=10$ bits of the shrunken
sequence $\{z_i\}=\{0,0,1,1,1,0,1,0,1,1\}$ and first rules
$R_1=R_2=0$. In fact, the 10 initial bits generate 8 bits at the
third cell in $\Delta 1$. These 8 bits are placed at the most left
extreme cell producing 6 new bits at cell 3 in $\Delta 2$. With
these 6 bits, we get 4 additional bits in $\Delta 3$. Finally, 2
new bits are obtained at cell 3 in the sub-triangle $\Delta 4$.
Since rules 90 and 150 are additive, the generated sub-sequences
will be sum of elements of the shrunken sequence. General
expressions can be deduced for the elements of any sub-sequence in
any chained sub-triangle. In fact, the \textit{i-th} sub-sequence
in the \textit{n-th} chained sub-triangle includes the bits $z_j$
corresponding to the exponents of $(P_{i-1}(x))^n$ where
$P_{i-1}(x)$ is the characteristic polynomial of the sub-automaton
$R_1R_2...R_{i-1}$, see equation (\ref{equation:0}). More
precisely, for the previous example the characteristic polynomial
of the sub-automaton $R_1R_2$ is $P_{2}(x)=x^2+1$. Then
$(P_{2}(x))^2=x^4+1$, $(P_{2}(x))^3=x^6+x^4+x^2+1$,
$(P_{2}(x))^4=x^8+1$, $\ldots$ Hence, $x_{3}^{t}$ in the different
sub-triangles will take the form:
\begin{center}
$
\begin{array}{c}
x_{3}^{t} = z_0+z_2 \;\;\;\; in \;\; \Delta 1 \\
x_{3}^{t} = z_0+z_4 \;\;\;\; in \;\; \Delta 2 \\
x_{3}^{t} = z_0+z_2+z_4+z_6 \;\;\;\;\; in \;\; \Delta 3 \\
x_{3}^{t} = z_0+z_8 \;\;\;\; in \;\; \Delta 4 \;\;\ldots\\
\end{array}
$
\end{center}
For the successive bits $x_{3}^{t+1}, x_{3}^{t+2},\ldots $ it
suffices to add $1$ to the previous subindexes. Table
\ref{table:headings3} shows the general expressions of the
sub-sequence elements in $\Delta 1$ and $\Delta 2$ for the example
under consideration.

\begin{table}
\begin{center}
\caption{Reconstruction of 4 chained sub-triangles from 10 bits of
the shrunken sequence} \label{table:headings2}
\begin{tabular}{ccccccccccc}
\hline\noalign{\smallskip}
$\Delta 1:$ & $R_1$ & $R_2$ & $R_3$ & \ldots & $\;\;\;\; \Delta 2:$  & $R_1$ & $R_2$ & $R_3$ & \ldots \\
\noalign{\smallskip} \hline
\hline
$\;$ & $\; 0 \;$ & $\; 0 \;$ & $\; 1 \;$ & \ldots & $\;$ & $\; 1 \;$ & $\; 1 \;$ & $\; 1 \;$ & \ldots \\
$\;$ & $\; 0 \;$ & $\; 1 \;$ & $\; 1 \;$ & \;\; & $\;$ & $\; 1 \;$
& $\; 0 \;$ & $\; 0 \;$ & \; \\
$\;$ & $\; 1 \;$ & $\; 1 \;$ & $\; 0 \;$ & \;\; & $\;$ & $\; 0 \;$ & $\; 1 \;$ & $\; 0 \;$ & \; \\
$\;$ & $\; 1 \;$ & $\; 1 \;$ & $\; 1 \;$ & \;\; & $\;$ & $\; 1 \;$ & $\; 0 \;$ & $\; 1 \;$ & \; \\
$\;$ & $\; 1 \;$ & $\; 0 \;$ & $\; 0 \;$ & \;\; & $\;$ & $\; 0 \;$ & $\; 0 \;$ & $\; 0 \;$ & \; \\
$\;$ & $\; 0 \;$ & $\; 1 \;$ & $\; 0 \;$ & \;\; & $\;$ & $\; 0 \;$ & $\; 0 \;$ & $\; 1 \;$ & \; \\
$\;$ & $\; 1 \;$ & $\; 0 \;$ & $\; 0 \;$ & \;\; & $\;$ & $\; 0 \;$ & $\; 1 \;$ & $\;  \;$ & \; \\
$\;$ & $\; 0 \;$ & $\; 1 \;$ & $\; 1 \;$ & \;\; & $\;$ & $\; 1 \;$ & $\;  \;$ & $\;  \;$ & \; \\
$\;$ & $\; 1 \;$ & $\; 1 \;$ & $\;  \;$ & \;\; & $\;$ & $\;  \;$ & $\;  \;$ & $\;  \;$ & \; \\
$\;$ & $\; 1 \;$ & $\;  \;$ & $\;  \;$ & \;\; & $\;$ & $\;  \;$ & $\;  \;$ & $\;  \;$ & \; \\
\noalign{\smallskip} \hline
$\Delta 3:$ & $R_1$ & $R_2$ & $R_3$ & \ldots & $\;\;\; \Delta 4:$  & $R_1$ & $R_2$ & $R_3$ & \ldots \\
\noalign{\smallskip} \hline \hline
$\;$ & $\; 1 \;$ & $\; 0 \;$ & $\; 1 \;$ & \ldots  & $\;$ & $\; 1 \;$ & $\; 1 \;$ & $\; 1 \;$ & \ldots  \\
$\;$ & $\; 0 \;$ & $\; 0 \;$ & $\; 1 \;$ & \;\; & $\;$ & $\; 1 \;$ & $\; 0 \;$ & $\; 1 \;$ & \; \\
$\;$ & $\; 0 \;$ & $\; 1 \;$ & $\; 0 \;$ & \;\; & $\;$ & $\; 0 \;$ & $\; 0 \;$ & $\;  \;$ & \; \\
$\;$ & $\; 1 \;$ & $\; 0 \;$ & $\; 0 \;$ & \;\; & $\;$ & $\; 0 \;$ & $\;  \;$  & $\;  \;$ & \; \\
$\;$ & $\; 0 \;$ & $\; 1 \;$ & $\;  \;$ & \;\;  & $\;$ & $\;  \;$  & $\;  \;$  & $\;  \;$ & \; \\
$\;$ & $\; 1 \;$ & $\;  \;$ & $\;  \;$ & \;\;   & $\;$ & $\;  \;$  & $\;  \;$  & $\;  \;$ & \; \\
\noalign{\smallskip} \hline
\end{tabular}
\end{center}
\end{table}

On the other hand, Lemmas (\ref{lemma:1}) and (\ref{lemma:2}) show
us that the shrunken sequence is the interleaving of
$2^{(L_{1}-1)}$ different shifts of an unique \textit{PN-}sequence
of length $2^{L_2}-1$ whose characteristic polynomial $P(x)$ is
given by equation (\ref{equation:5}). Consequently, the elements
of the shrunken sequence indexed $z_{d i}$, with $i\in\{0, 1,
\ldots , 2^{L_2}-2\}$ and $d=2^{(L_1-1)}$, belong to the same
\textit{PN-}sequence. Thus, if the element $x_{i}^t$ of the
\textit{i-th} sub-sequence in the \textit{n-th} chained
sub-triangle takes the general form:
\begin{equation}\label{equation:9}
x_{i}^t= z_{k_1}+z_{k_2}+ \ldots + z_{k_j}
\end{equation}
with
\begin{equation}\label{equation:10}
\;\;\;k_{l}\equiv 0 \;\;mod \;\; 2^{(L_1-1)} \;\;\; (l=1, \ldots ,
j),
\end{equation}
then $x_{i}^t$ can be rewritten as
\begin{equation}\label{equation:11}
x_{i}^t= z_{k_m},
\end{equation}
with $z_{k_m}$ satisfying equation (\ref{equation:10}). Therefore,
$\{x_{i}^{t}\}$, the \textit{i-th} sub-sequence in the
\textit{n-th} chained sub-triangle, is just a sub-sequence of the
shrunken sequence shifted a distance $\delta$ from the $r$ bits of
the intercepted sequence. The value of $\delta$ depends on the
extension field $GF(2^{L_{2}})$ generated by the roots of $P(x)$.
In brief, the chained sub-triangles enable us to reconstruct
additional bits of the shrunken sequence from bits of the
intercepted sequence.

\begin{table}
\begin{center}
\caption{General expressions for different sub-sequences in
$\Delta 1$ and $\Delta 2$ with $R_1=R_2=0$}
\label{table:headings3}
\begin{tabular}{ccccccccccc}
\hline\noalign{\smallskip}
$\Delta 1:$ & $R_1$ & $R_2$ & $R_3$ & \ldots & $\;\;\;\; \Delta 2: $  & $R_1$ & $R_2$ & $R_3$ & \ldots \\
\noalign{\smallskip} \hline \noalign{\smallskip} \hline
$\;$ & $\; z_0 \;$ & $\; z_1 \;$ & $z_0+z_2$ & \ldots & $\;$ & $z_0+z_2$ & $z_1+z_3$ & $z_0+z_4$ & \ldots \\
$\;$ & $z_1$ & $z_2$ & $z_1+z_3$ & \; & $\;$ & $z_1+z_3$ & $z_2+z_4$ & $z_1+z_5$ & \; \\
$\;$ & $z_2$ & $z_3$ & $z_2+z_4$ & \; & $\;$ & $z_2+z_4$ & $z_3+z_5$ & $z_2+z_6$ & \; \\
$\;$ & $z_3$ & $z_4$ & $z_3+z_5$ & \; & $\;$ & $z_3+z_5$ & $z_4+z_6$ & $z_3+z_7$ & \; \\
$\;$ & $z_4$ & $z_5$ & $z_4+z_6$ & \; & $\;$ & $z_4+z_6$ & $z_5+z_7$ & $z_4+z_8$ & \; \\
$\;$ & $z_5$ & $z_6$ & $z_5+z_7$ & \; & $\;$ & $z_5+z_7$ & $z_6+z_8$ & $z_5+z_9$ & \; \\
$\;$ & $z_6$ & $z_7$ & $z_6+z_8$ & \; & $\;$ & $z_6+z_8$ & $z_7+z_9$ & $\;$ & \; \\
$\;$ & $z_7$ & $z_8$ & $z_7+z_9$ & \; & $\;$ & $z_7+z_9$ & $\;$      & $\;$ & \; \\
$\;$ & $z_8$ & $z_9$ & $\;$      & \; & $\;$ & $\;$      & $\;$      & $\;$ & \; \\
$\;$ & $z_9$ & $\;$  & $\;$      & \; & $\;$ & $\;$      & $\;$      & $\;$ & \; \\
\noalign{\smallskip} \hline
\end{tabular}
\end{center}
\end{table}

\begin{table}[ht]
\caption{The shrunken sequence produced by the shrinking generator
described in subsection 5.3.} \label{table:headings4}
\renewcommand\arraystretch{1.5}
\renewcommand{\arraystretch}{0.93}
\noindent\[
\begin{tabular}{|c|c|cccccccc|}\hline
$\;$ & $\;$ & $C_1$ & $C_2$ & $C_3$ & $C_4$ & $C_5$ & $C_6$ &
$C_7$ & $C_8$ \\
\hline\noalign{\smallskip}\hline

$\;\;$ & $\;0\;$ & $\;1\;$ & $\;0\;$ & $\;1\;$ & $\;0\;$ & $\;0\;$ & $\;0\;$ & $\;0\;$ & $\;1\;$ \\
$\;\;$ & $\;1\;$ & $\;1\;$ & $\;0\;$ & $\;0\;$ & $\;1\;$ & $\;1\;$ & $\;1\;$ & $\;0\;$ & $\;0\;$ \\
$\;\;$ & $\;2\;$ & $\;1\;$ & $\;1\;$ & $\;0\;$ & $\;1\;$ & $\;0\;$ & $\;0\;$ & $\;1\;$ & $\;1\;$ \\
$\;\;$ & $\;3\;$ & $\;-\;$ & $\;-\;$ & $\;-\;$ & $\;-\;$ & $\;-\;$ & $\;-\;$ & $\;-\;$ & $\;-\;$ \\
$\;\;$ & $\;4\;$ & $\;-\;$ & $\;-\;$ & $\;-\;$ & $\;-\;$ & $\;-\;$ & $\;-\;$ & $\;-\;$ & $\;-\;$ \\
$\;\;$ & $\;5\;$ & $\;-\;$ & $\;-\;$ & $\;-\;$ & $\;-\;$ & $\;-\;$ & $\;-\;$ & $\;-\;$ & $\;-\;$ \\
$\;\;$ & $\;6\;$ & $\;-\;$ & $\;-\;$ & $\;-\;$ & $\;-\;$ & $\;-\;$ & $\;-\;$ & $\;-\;$ & $\;-\;$ \\
$\;\;$ & $\;7\;$ & $\;0\;$ & $\;1\;$ & $\;1\;$ & $\;1\;$ & $\;0\;$ & $\;0\;$ & $\;1\;$ & $\;0\;$ \\
$\;\;$ & $\;8\;$ & $\;-\;$ & $\;-\;$ & $\;-\;$ & $\;-\;$ & $\;-\;$ & $\;-\;$ & $\;-\;$ & $\;-\;$ \\
$\;\;$ & $\;9\;$ & $\;-\;$ & $\;-\;$ & $\;-\;$ & $\;-\;$ & $\;-\;$ & $\;-\;$ & $\;-\;$ & $\;-\;$ \\
$\;\;$ & $\;10\;$ & $\;-\;$ & $\;-\;$ & $\;-\;$ & $\;-\;$ & $\;-\;$ & $\;-\;$ & $\;-\;$ & $\;-\;$ \\
$\;\;$ & $\;11\;$ & $\;-\;$ & $\;-\;$ & $\;-\;$ & $\;-\;$ & $\;-\;$ & $\;-\;$ & $\;-\;$ & $\;-\;$ \\
$\;\;$ & $\;12\;$ & $\;-\;$ & $\;-\;$ & $\;-\;$ & $\;-\;$ & $\;-\;$ & $\;-\;$ & $\;-\;$ & $\;-\;$ \\
$\;\;$ & $\;13\;$ & $\;-\;$ & $\;-\;$ & $\;-\;$ & $\;-\;$ & $\;-\;$ & $\;-\;$ & $\;-\;$ & $\;-\;$ \\
$\;\;$ & $\;14\;$ & $\;-\;$ & $\;-\;$ & $\;-\;$ & $\;-\;$ & $\;-\;$ & $\;-\;$ & $\;-\;$ & $\;-\;$ \\
$\;\;$ & $\;15\;$ & $\;-\;$ & $\;-\;$ & $\;-\;$ & $\;-\;$ & $\;-\;$ & $\;-\;$ & $\;-\;$ & $\;-\;$ \\
$\;\;$ & $\;16\;$ & $\;-\;$ & $\;-\;$ & $\;-\;$ & $\;-\;$ & $\;-\;$ & $\;-\;$ & $\;-\;$ & $\;-\;$ \\
$\;\;$ & $\;17\;$ & $\;-\;$ & $\;-\;$ & $\;-\;$ & $\;-\;$ & $\;-\;$ & $\;-\;$ & $\;-\;$ & $\;-\;$ \\
$\;\;$ & $\;18\;$ & $\;-\;$ & $\;-\;$ & $\;-\;$ & $\;-\;$ & $\;-\;$ & $\;-\;$ & $\;-\;$ & $\;-\;$ \\
$\;\;$ & $\;19\;$ & $\;0\;$ & $\;0\;$ & $\;1\;$ & $\;1\;$ & $\;1\;$ & $\;1\;$ & $\;0\;$ & $\;1\;$ \\
$\;\;$ & $\;20\;$ & $\;0\;$ & $\;1\;$ & $\;0\;$ & $\;0\;$ & $\;1\;$ & $\;1\;$ & $\;1\;$ & $\;1\;$ \\
$\;\;$ & $\;21\;$ & $\;-\;$ & $\;-\;$ & $\;-\;$ & $\;-\;$ & $\;-\;$ & $\;-\;$ & $\;-\;$ & $\;-\;$ \\
$\;\;$ & $\;22\;$ & $\;-\;$ & $\;-\;$ & $\;-\;$ & $\;-\;$ & $\;-\;$ & $\;-\;$ & $\;-\;$ & $\;-\;$ \\
$\;b_4$ & $\;23\;$ & $\;1\;$ & $\;1\;$ & $\;1\;$ & $\;0\;$ & $\;1\;$ & $\;1\;$ & $\;1\;$ & $\;0\;$ \\
$\;\;$ & $\;24\;$ & $\;-\;$ & $\;-\;$ & $\;-\;$ & $\;-\;$ & $\;-\;$ & $\;-\;$ & $\;-\;$ & $\;-\;$ \\
$\;b_3$ & $\;25\;$ & $\;-\;$ & $\;-\;$ & $\;-\;$ & $\;-\;$ & $\;-\;$ & $\;-\;$ & $\;-\;$ & $\;-\;$ \\
$\;\;$ & $\;26\;$ & $\;-\;$ & $\;-\;$ & $\;-\;$ & $\;-\;$ & $\;-\;$ & $\;-\;$ & $\;-\;$ & $\;-\;$ \\
$\;b_2$ & $\;27\;$ & $\;-\;$ & $\;-\;$ & $\;-\;$ & $\;-\;$ & $\;-\;$ & $\;-\;$ & $\;-\;$ & $\;-\;$ \\
$\;\;$ & $\;28\;$ & $\;-\;$ & $\;-\;$ & $\;-\;$ & $\;-\;$ & $\;-\;$ & $\;-\;$ & $\;-\;$ & $\;-\;$ \\
$\;b_1$ & $\;29\;$ & $\;-\;$ & $\;-\;$ & $\;-\;$ & $\;-\;$ & $\;-\;$ & $\;-\;$ & $\;-\;$ & $\;-\;$ \\
$\;\;$ & $\;30\;$ & $\;-\;$ & $\;-\;$ & $\;-\;$ & $\;-\;$ & $\;-\;$ & $\;-\;$ & $\;-\;$ & $\;-\;$ \\
\hline
\end{tabular}
\]
\end{table}

The number of reconstructed bits depends on the amount of
intercepted bits. Indeed, if we know $N_l$ bits in each one of the
\textit{PN-}sequence shifts, then the total number of
reconstructed bits is given by:
\begin{equation}\label{equation:13}
\sum\limits_{l=1}^{2^{(L_1-1)}}\sum\limits_{k=2}^{N_{l}}\binom{N_{l}}{k}
\end{equation}
The required amount of intercepted sequence is $2^{L_1-1}$ that is
exponential in the length of the shorter register $SR_1$. Remark
that in this reconstruction process both reconstructed bits as
well as their positions on the shrunken sequence are known with
absolute certainty.

\subsection{Reconstruction of LFSR Initial States}
We denote by $IS_1=(a_0,a_1,$ $a_2,\ldots, a_{L_1-1})$ the initial
state of $SR_1$ and by $IS_2=(b_0,b_1,b_2,\ldots, b_{L_2-1})$ the
initial state of $SR_2$. In order to avoid ambiguities on the
initial states, it is assumed that $a_0=1$, thus the first element
of the shrunken sequence is $z_0=b_0$. In this way, the goal of
this attack is to determine the sub-vectors $(a_1,a_2,\ldots,
a_{L_1-1})$ as well as $(b_1,b_2,\ldots, b_{L_2-1})$.

According to equation (\ref{equation:1}), the period of the
shrunken sequence is $T=(2^{L_2}-1)\;2^{(L_1-1)}$, so that such a
sequence can be written as an $(2^{L_2}-1)\; \times \;
(2^{(L_1-1)})$ matrix whose elements are the bits of the shrunken
sequence. Its columns are denoted by $C_1, C_2, \ldots,
C_{2^{(L_1-1)}}$, respectively. Each column of the matrix is the
\textit{PN-}sequence above referenced starting at different
points. In addition, the first column $C_1$ corresponds to the
decimation of the sequence $\{b_{i}\}$ from $SR_2$ by a factor
$(2^{L_1}-1)$ \cite{Golomb}. Thus, we can compute the position of
the bits $b_1,b_2,\ldots, b_{L_2-1}$ on such a column. Indeed, the
\textit{i-th} bit, $b_i$, is at the $j_i-th$ position of $C_1$
where $j_i$ is solution of the equation:
\begin{equation}\label{equation:14}
j_{i}\;(2^{L_{1}}-1)\equiv i \;\;mod \;\; 2^{L_2}-1 \;\;\; (i=1,
\ldots , L_{2}-1).
\end{equation}
Moreover, the bits of $IS_1$ determine the initial bits of the
subsequent columns $C_i$ such as follows:
\begin{description}
\item [Hypothesis 1] If the first bits of $IS_1$ are
$(a_0=1,a_1=1)$, then $C_2$ will start at the $j_1-th$ position of
$C_1$ given by equation (\ref{equation:14}). \item [Hypothesis 2]
If the first bits of $IS_1$ are $(a_0=1,a_1=0,a_2=1)$, then $C_2$
will start at the $j_2-th$ position of $C_1$ given by equation
(\ref{equation:14}). \end{description}
\hspace{6.5cm} $\vdots$ \\
\begin{description}\item [Hypothesis n] If the first bits of $IS_1$ are
$(a_0=1,a_1=0,\ldots, a_{n-1}=0,a_n=1)$, then $C_2$ will start at
the $j_n-th$ position of $C_1$ given by equation
(\ref{equation:14}).
\end{description}

We can formulate different hypothesis covering the first bits of
$IS_1$ as well as each new hypothesis determines the initial bit
of the following column. As we have intercepted and reconstructed
bits in the columns $C_i$, we can check the previous hypothesis
until getting a contradiction. In that case, all the $IS_1$
starting with the wrong configuration must be rejected. The search
continues through the configurations of $a_i$ free of
contradiction by formulating new hypothesis. In brief, the
attacker has not to traverse an entire search tree including all
the initial states of $SR_1$, but the search is concentrated
exclusively on the configurations not exhibiting contradiction
with regard to the available bits. In this sense, the proposed
attack reduces considerably the exhaustive search over the initial
states of $SR_1$ as many contradictions occur at the first levels
of the tree. On the other hand, the bits of the register $SR_2$
are easily determined as the starting bits of $C_2, C_3, C_4,
\ldots $ in each one of the non-rejected branches. An illustrative
example of Phases 1 and 2 is presented in the next subsection.

\subsection{An Illustrative Example}
Let us consider a shrinking generator with the following
parameters: $L_1=4$, $L_2=5$, $C_{1}(x)=1+x^{3}+x^{4}$ and
$C_{2}(x)=1+x+x^{3}+x^{4}+x^{5}$. According to equation
 (\ref{equation:5}), we can compute the polynomial
$P(x)=1+x+x^{2}+x^{4}+x^{5}$ while the two basic automata $1 \; 0
\; 0 \; 0 \; 0 $ and $ 0 \; 0 \; 0 \; 0 \; 1$ are obtained from
the algorithm of Cattell and Muzio. The corresponding CA of length
$L=40$ are computed via the algorithm developed in section 3.
Indeed, they are $CA_1=0060110600$ and $CA_2=8C0300C031$ in
hexadecimal notation. In addition, let $\alpha$ be a root of
$P(x)$ that is $\alpha^5=\alpha^4+\alpha^2+\alpha+1$ as well as a
generator element of the extension field $GF(2^{L_{2}})$. The
period of the shrunken sequence is
$T=(2^{L_2}-1)\cdot2^{(L_1-1)}=248$ and the number of interleaved
\textit{PN-}sequences is $2^{(L_1-1)}=8$. Finally, the intercepted
sequence of length $r=24$ is: $\{z_0,z_1, \ldots,
z_{23}\}=$$\{1,0,1,0,0,0,0,1,1,0,0,1,1,1,0,0,1,1,0,1,0,0,1,1\}$.
With the previous premises, we accomplish Phases 1 and 2.

\textbf{Phase 1:}
\begin{description}
\item [For $CA_1$] The chained sub-triangles provide the following
reconstructed bits. For $i=3$, sub-automaton $R_1R_2$ and
$P_2(x)=x^2+1$.
\begin{itemize}
\item  In $\Delta 4$, $x_{3}^{t} = z_0+z_8$, $x_{3}^{t+1} =
z_1+z_9, \;\ldots\;, x_{3}^{t+15} = z_{15}+z_{23}$. Considering
$z_0,z_8$ as the first and second element of the
\textit{PN-}sequence and keeping in mind that in $GF(2^{L_{2}})$
the equality $1+\alpha=\alpha^{19}$ holds, we get $x_{3}^{t} =
z_{19 \cdot 8}=z_{152}$, $x_{3}^{t+1} = z_{153},\;\ldots\;,
x_{3}^{t+15} = z_{167}$. Thus, 16 new bits of the shrunken
sequence have been reconstructed at positions $152, 153, \ldots,
167$. \item In $\Delta 8$, $x_{3}^{t} = z_0+z_{16}$, $x_{3}^{t+1}
= z_1+z_{17}, \;\ldots\;, x_{3}^{t+7} = z_{7}+z_{23}$. As
$1+\alpha^2=\alpha^{7}$, we get $x_{3}^{t} = z_{7 \cdot
8}=z_{56}$, $x_{3}^{t+1} = z_{57},\;\ldots\;, x_{3}^{t+7} =
z_{63}$. Thus, 8 new bits of the shrunken sequence have been
reconstructed at positions $56, 57, \ldots, 63$.
\end{itemize}
 \item
[For $CA_2$] The chained sub-triangles provide the following
reconstructed bits. For $i=3$, sub-automaton $R_1R_2$ and
$P_2(x)=x^2+x+1$.
\begin{itemize}
\item  In $\Delta 8$, $x_{3}^{t} = z_0+z_8+z_{16}$, $x_{3}^{t+1} =
z_1+z_{9}+z_{17}, \;\ldots\;, x_{3}^{t+7} = z_7+z_{15}+z_{23}$. As
$1+\alpha+\alpha^2=\alpha^{23}$, we get $x_{3}^{t} = z_{23 \cdot
8}=z_{184}$, $x_{3}^{t+1} = z_{185},\;\ldots\;, x_{3}^{t+7} =
z_{191}$. Thus, 8 new bits of the shrunken sequence have been
reconstructed at positions $184, 185, \ldots, 191$.
\end{itemize}
\end{description}
After Phase 1, the known bits of the shrunken sequence are
depicted in Table \ref{table:headings4}. Rows 0,1,2 correspond to
intercepted bits while rows 7, 19, 20 and 23 correspond to
reconstructed bits. The symbol $-$ represents the unknown bits. In
brief, from 24 intercepted bits a total of 32 bits have been
reconstructed.

\textbf{Phase2:} According to equation (\ref{equation:14}), the
bits $b_1, b_2, b_3, b_4$ are placed at positions $29, 27,$  $25,
23$ of column $C_1$, respectively (see the first column of Table
\ref{table:headings4}). On the other hand, Table 5 shows the
sequences corresponding to the following hypothesis.
\begin{description}
\item [Hypothesis 1] If the first bits of $IS_1$ are
$(a_0=1,a_1=1)$, then $C_2$ will start at the $29^{th}$ position
of $C_1$ given rise to the column $H_1$. In row 2, $H_1$ and $C_2$
have a common bit without contradiction. The union of both
sequences allows us to construct $C_{2}^{1}$ the second column of
the matrix for this hypothesis. A total of 13 bits are then known
in $C_{2}^{1}$. \item [Hypothesis 2] If the first bits of $IS_1$
are $(a_0=1,a_1=0,a_2=1)$, then $C_2$ will start at the $27^{th}$
position of $C_1$ given rise to the column $H_2$. In row 23, $H_2$
and $C_2$ have a common bit with contradiction (starred bits).
Thus, the initial states of $SR_1$ starting with bits $101$ must
be rejected. \item [Hypothesis 3] If the first bits of $IS_1$ are
$(a_0=1,a_1=0,a_2=0,a_3=1)$, then $C_2$ will start at the
$25^{th}$ position of $C_1$ given rise to the column $H_3$. In row
7, $H_3$ and $C_2$ have a common bit without contradiction. The
union of both sequences allows us to construct $C_{2}^{3}$ the
second column of the matrix for this hypothesis. A total of 13
bits are then known in $C_{2}^{3}$. \item [Hypothesis 4] If the
first bits of $IS_1$ are $(a_0=1,a_1=0,a_2=0,a_3=0,a_4=1)$, then
$C_2$ will start at the $23^{th}$ position of $C_1$ given rise to
the column $H_4$. In row 0, $H_2$ and $C_2$ have a common bit with
contradiction (starred bits). Thus, the initial state of $SR_1$
$1000$ must be rejected.
\end{description}

\begin{table}[ht]
\caption{Different hypothesis formulated on the bits of $SR_1$}
\label{table:headings5}
\renewcommand{\arraystretch}{0.75}
\noindent\[
\begin{tabular}{|c|cccc||ccc||cccc||ccc|}

\hline $\;$ & $C_1$ & $H_1$ & $C_2$ & $C_{2}^{1}$ & $C_1$ & $H_2$
& $C_2$ & $C_1$ & $H_3$ & $C_2$ & $C_{2}^{3}$ &
$C_1$ & $H_4$ & $C_2$ \\
\hline\noalign{\smallskip}\hline
$0$ & $1$ & $-$ & $0$ & $0$ & $1$ & $-$ & $0$ & $1$ & $-$ & $0$ & $0$ & $1$ & $\;\;1^*$ & $\;\;0^*$ \\
$1$ & $1$ & $-$ & $0$ & $0$ & $1$ & $-$ & $0$ & $1$ & $-$ & $0$ & $0$ & $1$ & $-$ & $0$ \\
$2$ & $1$ & $1$ & $1$ & $1$ & $1$ & $-$ & $1$ & $1$ & $-$ & $1$ & $1$ & $1$ & $-$ & $1$ \\
$3$ & $-$ & $1$ & $-$ & $1$ & $-$ & $-$ & $-$ & $-$ & $-$ & $-$ & $-$ & $-$ & $-$ & $-$ \\
$4$ & $-$ & $1$ & $-$ & $1$ & $-$ & $1$ & $-$ & $-$ & $-$ & $-$ & $-$ & $-$ & $-$ & $-$ \\
$5$ & $-$ & $-$ & $-$ & $-$ & $-$ & $1$ & $-$ & $-$ & $-$ & $-$ & $-$ & $-$ & $-$ & $-$ \\
$6$ & $-$ & $-$ & $-$ & $-$ & $-$ & $1$ & $-$ & $-$ & $1$ & $-$ & $1$ & $-$ & $-$ & $-$ \\
$7$ & $0$ & $-$ & $1$ & $1$ & $0$ & $-$ & $1$ & $0$ & $1$ & $1$ & $1$ & $0$ & $-$ & $1$ \\
$8$ & $-$ & $-$ & $-$ & $-$ & $-$ & $-$ & $-$ & $-$ & $1$ & $-$ & $1$ & $-$ & $1$ & $-$ \\
$9$ & $-$ & $0$ & $-$ & $0$ & $-$ & $-$ & $-$ & $-$ & $-$ & $-$ & $-$ & $-$ & $1$ & $-$ \\
$10$ & $-$ & $-$ & $-$ & $-$ & $-$ & $-$ & $-$ & $-$ & $-$ & $-$ & $-$ & $-$ & $1$ & $-$ \\
$11$ & $-$ & $-$ & $-$ & $-$ & $-$ & $0$ & $-$ & $-$ & $-$ & $-$ & $-$ & $-$ & $-$ & $-$ \\
$12$ & $-$ & $-$ & $-$ & $-$ & $-$ & $-$ & $-$ & $-$ & $-$ & $-$ & $-$ & $-$ & $-$ & $-$ \\
$13$ & $-$ & $-$ & $-$ & $-$ & $-$ & $-$ & $-$ & $-$ & $0$ & $-$ & $0$ & $-$ & $-$ & $-$ \\
$14$ & $-$ & $-$ & $-$ & $-$ & $-$ & $-$ & $-$ & $-$ & $-$ & $-$ & $-$ & $-$ & $-$ & $-$ \\
$15$ & $-$ & $-$ & $-$ & $-$ & $-$ & $-$ & $-$ & $-$ & $-$ & $-$ & $-$ & $-$ & $0$ & $-$ \\
$16$ & $-$ & $-$ & $-$ & $-$ & $-$ & $-$ & $-$ & $-$ & $-$ & $-$ & $-$ & $-$ & $-$ & $-$ \\
$17$ & $-$ & $-$ & $-$ & $-$ & $-$ & $-$ & $-$ & $-$ & $-$ & $-$ & $-$ & $-$ & $-$ & $-$ \\
$18$ & $-$ & $-$ & $-$ & $-$ & $-$ & $-$ & $-$ & $-$ & $-$ & $-$ & $-$ & $-$ & $-$ & $-$ \\
$19$ & $0$ & $-$ & $0$ & $0$ & $0$ & $-$ & $0$ & $0$ & $-$ & $0$ & $0$ & $0$ & $-$ & $0$ \\
$20$ & $0$ & $-$ & $1$ & $1$ & $0$ & $-$ & $1$ & $0$ & $-$ & $1$ & $1$ & $0$ & $-$ & $1$ \\
$21$ & $-$ & $0$ & $-$ & $0$ & $-$ & $-$ & $-$ & $-$ & $-$ & $-$ & $-$ & $-$ & $-$ & $-$ \\
$22$ & $-$ & $0$ & $-$ & $0$ & $-$ & $-$ & $-$ & $-$ & $-$ & $-$ & $-$ & $-$ & $-$ & $-$ \\
$23$ & $1$ & $-$ & $1$ & $1$ & $1$ & $\;\;0^*$ & $\;\;1^*$ & $1$ & $-$ & $1$ & $1$ & $1$ & $-$ & $1$ \\
$24$ & $-$ & $-$ & $-$ & $-$ & $-$ & $0$ & $-$ & $-$ & $-$ & $-$ & $-$ & $-$ & $-$ & $-$ \\
$25$ & $-$ & $1$ & $-$ & $1$ & $-$ & $-$ & $-$ & $-$ & $0$ & $-$ & $0$ & $-$ & $-$ & $-$ \\
$26$ & $-$ & $-$ & $-$ & $-$ & $-$ & $-$ & $-$ & $-$ & $0$ & $-$ & $0$ & $-$ & $-$ & $-$ \\
$27$ & $-$ & $-$ & $-$ & $-$ & $-$ & $1$ & $-$ & $-$ & $-$ & $-$ & $-$ & $-$ & $0$ & $-$ \\
$28$ & $-$ & $-$ & $-$ & $-$ & $-$ & $-$ & $-$ & $-$ & $-$ & $-$ & $-$ & $-$ & $0$ & $-$ \\
$29$ & $-$ & $-$ & $-$ & $-$ & $-$ & $-$ & $-$ & $-$ & $1$ & $-$ & $1$ & $-$ & $-$ & $-$ \\
$30$ & $-$ & $-$ & $-$ & $-$ & $-$ & $-$ & $-$ & $-$ & $-$ & $-$ & $-$ & $-$ & $-$ & $-$ \\
\hline \hline \multicolumn{1}{|c|}{$\;$} &
\multicolumn{4}{c||}{Hypothesis 1} &
\multicolumn{3}{c||}{Hypothesis 2} &
\multicolumn{4}{c||}{Hypothesis 3} &
\multicolumn{3}{c|}{Hypothesis 4} \\
\hline
\end{tabular}
\]
\end{table}

\begin{table}[ht]
\caption{Different hypothesis formulated on the bits of $SR_1$}
\label{table:headings6}
\renewcommand{\arraystretch}{0.75}
\noindent\[
\begin{tabular}{|c||ccc||ccc||cc|}
\hline $\;$ & $C_1$ & $H_5$ & $C_3$ & $C_1$ & $H_6$ & $C_3$ & $C_1$ & $C_{2}^{3}$\\
\hline\noalign{\smallskip}\hline
$0$  & $1$ & $-$ & $1$ & $1$ & $1$ & $1$ & $1$ & $0$ \\
$1$  & $1$ & $-$ & $0$ & $1$ & $0$ & $0$ & $1$ & $0$ \\
$2$  & $1$ & $-$ & $0$ & $1$ & $\;\;1^*$ & $\;\;0^*$ & $1$ & $1$ \\
$3$  & $-$ & $-$ & $-$ & $-$ & $-$ & $-$ & $-$ & $-$ \\
$4$  & $-$ & $1$ & $-$ & $-$ & $0$ & $-$ & $-$ & $-$ \\
$5$  & $-$ & $1$ & $-$ & $-$ & $0$ & $-$ & $-$ & $-$ \\
$6$  & $-$ & $1$ & $-$ & $-$ & $0$ & $-$ & $-$ & $1$ \\
$7$  & $0$ & $-$ & $1$ & $0$ & $-$ & $1$ & $0$ & $1$ \\
$8$  & $-$ & $-$ & $-$ & $-$ & $1$ & $-$ & $-$ & $1$ \\
$9$  & $-$ & $-$ & $-$ & $-$ & $1$ & $-$ & $-$ & $-$ \\
$10$ & $-$ & $-$ & $-$ & $-$ & $1$ & $-$ & $-$ & $-$ \\
$11$ & $-$ & $0$ & $-$ & $-$ & $-$ & $-$ & $-$ & $-$ \\
$12$ & $-$ & $-$ & $-$ & $-$ & $-$ & $-$ & $-$ & $-$ \\
$13$ & $-$ & $-$ & $-$ & $-$ & $-$ & $-$ & $0$ & $0$ \\
$14$ & $-$ & $-$ & $-$ & $-$ & $-$ & $-$ & $1$ & $-$ \\
$15$ & $-$ & $-$ & $-$ & $-$ & $0$ & $-$ & $-$ & $-$ \\
$16$ & $-$ & $-$ & $-$ & $-$ & $-$ & $-$ & $-$ & $-$ \\
$17$ & $-$ & $-$ & $-$ & $-$ & $-$ & $-$ & $1$ & $-$ \\
$18$ & $-$ & $-$ & $-$ & $-$ & $-$ & $-$ & $-$ & $-$ \\
$19$ & $0$ & $-$ & $1$ & $0$ & $-$ & $1$ & $0$ & $0$ \\
$20$ & $0$ & $-$ & $0$ & $0$ & $-$ & $0$ & $0$ & $1$ \\
$21$ & $-$ & $-$ & $-$ & $-$ & $-$ & $-$ & $-$ & $-$ \\
$22$ & $-$ & $-$ & $-$ & $-$ & $-$ & $-$ & $-$ & $-$ \\
$23$ & $1$ & $\;\;0^*$ & $\;\;1^*$ & $1$ & $-$ & $1$ & $1$ & $1$ \\
$24$ & $-$ & $0$ & $-$ & $0$ & $-$ & $-$ & $-$ & $-$ \\
$25$ & $-$ & $-$ & $-$ & $1$ & $-$ & $-$ & $0$ & $0$ \\
$26$ & $-$ & $-$ & $-$ & $-$ & $-$ & $-$ & $0$ & $0$ \\
$27$ & $-$ & $1$ & $-$ & $-$ & $0$ & $-$ & $1$ & $-$ \\
$28$ & $-$ & $-$ & $-$ & $-$ & $0$ & $-$ & $-$ & $-$ \\
$29$ & $-$ & $-$ & $-$ & $-$ & $-$ & $-$ & $-$ & $1$ \\
$30$ & $-$ & $-$ & $-$ & $-$ & $-$ & $-$ & $-$ & $-$ \\
\hline \noalign{\smallskip} \hline \multicolumn{1}{|c||}{$\;$} &
\multicolumn{3}{c||}{Hypothesis 5} &
\multicolumn{3}{c||}{Hypothesis 6}& \multicolumn{2}{c|}{Solution} \\
\hline
\end{tabular}
\]
\end{table}

\begin{figure}
\begin{center}
\includegraphics[bb= 10 12 320 200, scale=0.85]{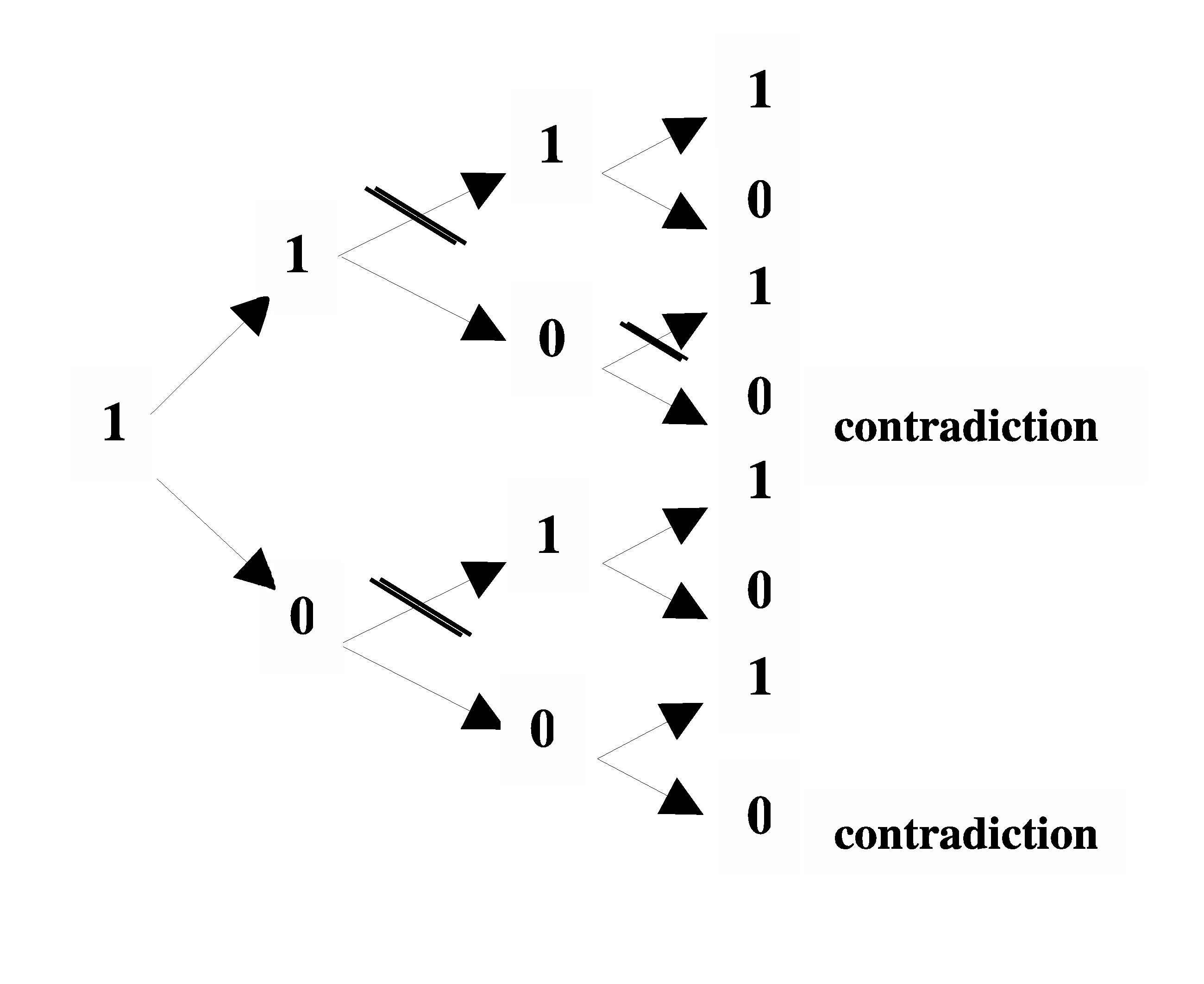}

\caption{Search tree for the initial sates of $SR_1$}
\label{figure:headings1}
\end{center}
\end{figure}

On the hypothesis free of contradiction, we can formulate other
ones depicted in Table \ref{table:headings6}
\begin{description}
\item [Hypothesis 5] If the first bits of $IS_1$ are
$(a_0=1,a_1=1, a_2=1)$, then $C_3$ will start at the $27^{th}$
position of $C_1$ given rise to the column $H_5$. In row 23, $H_5$
and $C_3$ have a common bit with contradiction (starred bits).
Thus, the initial states of $SR_1$ starting with bits $111$ must
be rejected. \item [Hypothesis 6] If the first bits of $IS_1$ are
$(a_0=1,a_1=1,a_2=0, a_3=0,a_4=1)$, then $C_3$ will start at the
$23^{th}$ position of $C_1$ given rise to the column $H_6$. Bits
$24$ and $25$ of $C_1$ have been deduced from $C_{2}^{1}$ in
Hypothesis 1. In row 2, $H_6$ and $C_6$ have a common bit with
contradiction (starred bits). Thus, the initial state of $SR_1$
$1100$ must be rejected.
\end{description}
From Hypothesis 5 and 6, Hypothesis 1 must be rejected. Remark
that the configuration $(a_0=1,a_1=0,a_2=0,a_3=1)$ in Hypothesis 3
is the only one free of contradiction. See the search tree in
figure \ref{figure:headings1}. Thus, it corresponds to the actual
initial state of $SR_1$. The successive bits of $SR_1$, that is
the \textit{PN-}sequence $\{1,0,0,1,0,0,0,1, \ldots \}$, are
checked by the successive columns $C_4, C_5, \ldots , C_8 $ of the
shrunken sequence. Concerning the initial state of $SR_2$, in
Table \ref{table:headings6} (column Solution) we can see that bits
$b_4, b_3, b_2$ can be obtained from the known bits of $C_1$ in
rows 23, 25 and 27 respectively. In fact, $b_4=1, b_3=0, b_2=1$.
The bit $b_1$ in row 29 satisfies the equality
\begin{equation}\label{equation:15}
b_1=z_{29 \cdot 8}=z_{1 \cdot 8}+z_{2 \cdot 8}+z_{4 \cdot 8},
\end{equation}
as $\alpha+\alpha^2+\alpha^4=\alpha^{29}$ in the extension field
$GF(2^{L_{2}})$. We know that $z_{8}=1, z_{16}=1$ while $z_{32}$
can be easily deduced from the equality $z_{14 \cdot 8}=z_{1 \cdot
8}+z_{4 \cdot 8}$ as $1+\alpha^4=\alpha^{14}$. Thus,
$z_{32}=1+1=0$ and substituting in $b_1$ we get $b_1=1+1+0=0$.

The final issues of Phases 1 and 2 are the initial states of both
LFSRs $IS_1=(a_0,a_1,\ldots, a_{3})=(1,0,0,1)$ and
$IS_2=(b_0,b_1,\ldots, b_{4})=(1,0,1,0,1)$. From the knowledge of
both initial states the whole shrunken sequence can be
reconstructed.

\subsection{Computational Features of the Attack}
The computational complexity of the previous cryptanalytic attack
can be considered in two different phases: off-line and on-line
complexity.

\textit{Off-line computational complexity:} This phase is to be
executed before intercepting sequence. It includes:
\begin{itemize}
\item Computation of the characteristic polynomials $P_{i}(x)$ of
the sub-automata $R_1R_2\ldots R_i$ $\;(1<i \leq l)$ by means of
equation (\ref{equation:0}) where $l$ is related to the amount of
intercepted sequence ($l \cdot 2^{L_{1}-1}\sim r$). This
computation is necessary in order to obtain general expressions
for the elements of the chained sub-triangles in the
reconstruction procedure. \item Computation of the positions of
the bits $b_i$ $(i=1,2,\ldots,L_2-1)$ on $C_1$ the first column of
the shrunken sequence matrix by means of equation
(\ref{equation:14}). This computation is necessary in order to
determine the bits of the initial state of $SR_2$. \item
Computation of different elements of the extension field
$GF(2^{L_{2}})$ such as $1+\alpha$, $1+\alpha^2,\ldots ,
1+\alpha^{N_l}$ and linear combinations of them by means of the
Zech log table method \cite{Assis} for arithmetic over
$GF(2^{m})$. This computation is necessary in order to determine
the distance between the intercepted sequence and the portions of
reconstructed shrunken sequence.
\end{itemize}

\textit{On-line computational complexity:} This phase is to be
executed after intercepting sequence. According to the previous
subsections, the computational method consists in the comparison
of series of bits coming from formulated hypothesis and from
intercepted/reconstructed bits. The comparison is realized by
means of bit-wise logical operations so the computational
complexity is rather low. Occasionally, the computation of the any
element of $GF(2^{L_{2}})$ must be realized in order to determine
additional elements of the \textit{PN-}sequences. The most
consuming time of this cryptanalytic attack is the search over the
$2^{L_1-1}$ possible initial states of $SR_1$ (supposed $a_0=1$).
Due to contradictions found in the first levels of the search
tree, the exhaustive search can be dramatically improved. On
average, we can say that in the worst case the search can be
reduced to the half, so that the computational complexity of this
attack is $O(2^{L_{1}-2})$. In addition, several considerations
must be kept in mind:
\begin{enumerate}
\item The improved exhaustive search is carried out over the state
space of the shorter register $SR_1$. \item Every checking of
hypothesis is realized only over the $1's$ of the configuration
under consideration, then the procedure speeds for configurations
with a low number of $1's$.
\end{enumerate}

Finally, comparing the proposed attack with those ones found in
the literature we get that all of them are exponential in the
lengths of the registers. In particular, the complexity of the
divide-and-conquer attack proposed in \cite{Simpson} is
$O(2^{L_{1}})$. The probabilistic correlation attack described in
\cite{Golic} has a computational complexity of $O(L_2^2 \cdot
2^{L_{2}})$. Also the probabilistic correlation attack introduced
in \cite{Johansson} is exponential in $L_2$. In this work a
deterministic attack has been proposed that improves the
complexity of the previous cryptanalytic approaches.

\section{Conclusions}

A wide family of LFSR-\;based sequence generators, the so-\;called
Clock Controlled Shrinking Generators, has been analyzed and
identified with a subset of linear cellular automata. In this way,
sequence generators conceived and designed as complex nonlinear
models can be written in terms of simple linear models. An easy
algorithm to compute the pair of one-\;dimensional linear hybrid
cellular automata that generate the CCSG output sequences has been
derived. In addition, a cryptanalytic attack that reconstructs the
output sequence of such generators has been proposed too. The
procedure is based on the linearity of these CA-based models as
well as on the characteristics of this class of generators. The
cryptanalytic approach is deterministic and improves an exhaustive
search over the states of the shorter register. Computing the
initial state of the longer register is a direct consequence of
the previous step. From the obtained results, we can create linear
cellular automata-\;based models to analyze/cryptanalyze the class
of clock-\;controlled generators.

\bibliographystyle{amsalpha}

\end{document}